\documentclass[aps,pra,noshowpacs,twocolumn,preprintnumbers,superscriptaddress,floatfix]{revtex4}
\usepackage{amsmath,amssymb}
\usepackage{multirow}
\usepackage{graphicx}
\usepackage{epsfig}
\usepackage{times}
\usepackage{pdfpages}
\usepackage{url}
\usepackage[colorlinks,citecolor=black,linkcolor=black,urlcolor=blue,breaklinks=True]{hyperref}
\usepackage{yhmath}

\begin{document}

\title{A Scalable Photonic Computer Solving the Subset Sum Problem}

\author{Xiao-Yun Xu}
\affiliation{Center for Integrated Quantum Information Technologies (IQIT), School of Physics and Astronomy and State Key Laboratory of Advanced Optical Communication Systems and Networks, Shanghai Jiao Tong University, Shanghai 200240, China}
\affiliation{CAS Center for Excellence and Synergetic Innovation Center in Quantum Information and Quantum Physics, University of Science and Technology of China, Hefei, Anhui 230026, China}

\author{Xuan-Lun Huang}
\affiliation{Center for Integrated Quantum Information Technologies (IQIT), School of Physics and Astronomy and State Key Laboratory of Advanced Optical Communication Systems and Networks, Shanghai Jiao Tong University, Shanghai 200240, China}
\affiliation{CAS Center for Excellence and Synergetic Innovation Center in Quantum Information and Quantum Physics, University of Science and Technology of China, Hefei, Anhui 230026, China}

\author{Zhan-Ming Li}
\affiliation{Center for Integrated Quantum Information Technologies (IQIT), School of Physics and Astronomy and State Key Laboratory of Advanced Optical Communication Systems and Networks, Shanghai Jiao Tong University, Shanghai 200240, China}
\affiliation{CAS Center for Excellence and Synergetic Innovation Center in Quantum Information and Quantum Physics, University of Science and Technology of China, Hefei, Anhui 230026, China}

\author{Jun Gao}
\affiliation{Center for Integrated Quantum Information Technologies (IQIT), School of Physics and Astronomy and State Key Laboratory of Advanced Optical Communication Systems and Networks, Shanghai Jiao Tong University, Shanghai 200240, China}
\affiliation{CAS Center for Excellence and Synergetic Innovation Center in Quantum Information and Quantum Physics, University of Science and Technology of China, Hefei, Anhui 230026, China}

\author{Zhi-Qiang Jiao}
\affiliation{Center for Integrated Quantum Information Technologies (IQIT), School of Physics and Astronomy and State Key Laboratory of Advanced Optical Communication Systems and Networks, Shanghai Jiao Tong University, Shanghai 200240, China}
\affiliation{CAS Center for Excellence and Synergetic Innovation Center in Quantum Information and Quantum Physics, University of Science and Technology of China, Hefei, Anhui 230026, China}

\author{Yao Wang}
\affiliation{Center for Integrated Quantum Information Technologies (IQIT), School of Physics and Astronomy and State Key Laboratory of Advanced Optical Communication Systems and Networks, Shanghai Jiao Tong University, Shanghai 200240, China}
\affiliation{CAS Center for Excellence and Synergetic Innovation Center in Quantum Information and Quantum Physics, University of Science and Technology of China, Hefei, Anhui 230026, China}

\author{Ruo-Jing Ren}
\affiliation{Center for Integrated Quantum Information Technologies (IQIT), School of Physics and Astronomy and State Key Laboratory of Advanced Optical Communication Systems and Networks, Shanghai Jiao Tong University, Shanghai 200240, China}
\affiliation{CAS Center for Excellence and Synergetic Innovation Center in Quantum Information and Quantum Physics, University of Science and Technology of China, Hefei, Anhui 230026, China}

\author{H. P. Zhang}
\affiliation{School of Physics and Astronomy, Institute of Natural Sciences, Shanghai Jiao Tong University, Shanghai 200240, China.}

\author{Xian-Min Jin}
\thanks{xianmin.jin@sjtu.edu.cn}
\affiliation{Center for Integrated Quantum Information Technologies (IQIT), School of Physics and Astronomy and State Key Laboratory of Advanced Optical Communication Systems and Networks, Shanghai Jiao Tong University, Shanghai 200240, China}
\affiliation{CAS Center for Excellence and Synergetic Innovation Center in Quantum Information and Quantum Physics, University of Science and Technology of China, Hefei, Anhui 230026, China}
\affiliation{Institute for Quantum Science and Engineering and Department of Physics, Southern University of Science and Technology, Shenzhen 518055, China}

\maketitle


\textbf{The subset sum problem is a typical NP-complete problem that is hard to solve efficiently in time due to the intrinsic superpolynomial-scaling property. Increasing the problem size results in a vast amount of time consuming in conventionally available computers. Photons possess the unique features of  extremely high propagation speed, weak interaction with environment and low detectable energy level, therefore can be a promising candidate to meet the challenge by constructing an a photonic computer computer. However, most of optical computing schemes, like Fourier transformation, require very high operation precision and are hard to scale up. Here, we present a chip built-in photonic computer to efficiently solve the subset sum problem. We successfully map the problem into a waveguide network in three dimensions by using femtosecond laser direct writing technique. We show that the photons are able to sufficiently  dissipate into the networks and search all the possible paths for solutions in parallel. In the case of successive primes the proposed approach exhibits a dominant superiority in time consumption even compared with supercomputers. Our results confirm the ability of light to realize a complicated computational function that is intractable with conventional computers, and suggest the subset sum problem as a good benchmarking platform for the race between photonic and conventional computers on the way towards ``photonic supremacy".
}

\section*{\large Introduction}
NP-complete problems \cite{Garey_1979_computers} are typically defined as the problems solvable in polynomial time on a non-deterministic Turing machine (NTM), which indicates such problems are computationally hard on conventional electronic computers, a general type of deterministic Turing machines. The subset sum problem (SSP) with practical application in resource allocation \cite{Darmann_2014_euroopean} is a benchmark NP-complete problem \cite{Karp_1972_complexity} and its intractability has been harnessed in cryptosystems resistant to quantum attacks \cite{Okamoto_2000_advances,Kate_2011_security}. Given a finite set $S$ of $N$ integers, the SSP asks whether there is a subset of $S$ whose sum is equal to the target $T$. Apparently, the number of subset grows exponentially with the problem size $N$, which leads to an exponential time scaling and thus strongly limits the size of the problem that can be tackled in reality.

\begin{figure*}
\centering
\includegraphics[width=2 \columnwidth]{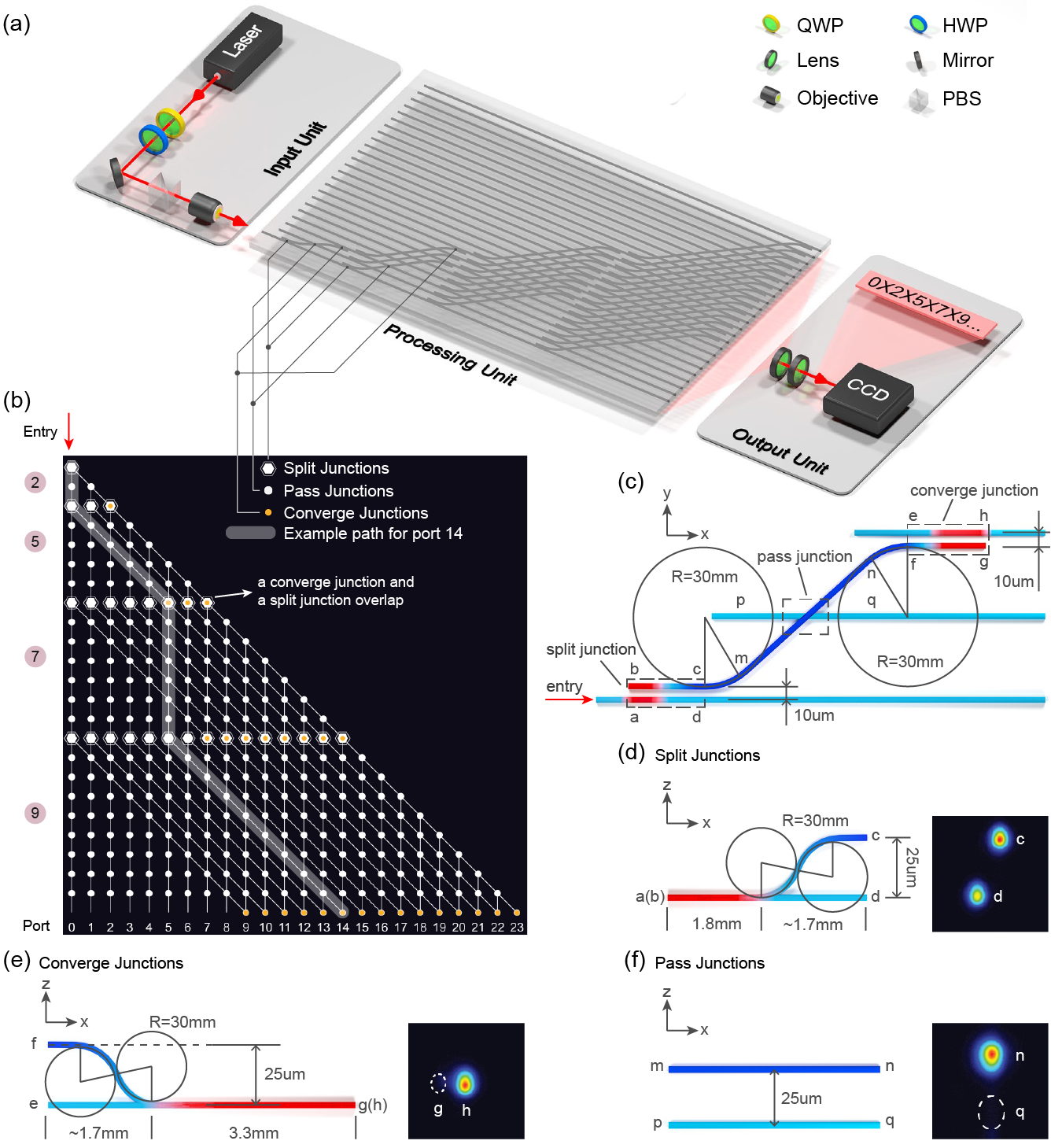}
\caption{\textbf{Schematic of the design and setup.} \textbf{(a)} A power-adjustable and horizontally polarized optical source is guaranteed by the quarter-wave plate (QWP), half-wave plate (HWP) and polarization beam splitter (PBS) in the input unit. The photons at 810 nm are prepared and coupled into the network in the processing unit, then travel to generate all possible subset sums. The evolution results at the output ports are retrieved by the CCD to testify the existence of the corresponding sums. \textbf{(b)} The abstract network for the specific instance $\{2,5,7,9\}$ is composed of three different kinds of nodes representing split junctions, pass junctions and converge junctions, respectively. \textbf{Split junctions} (hexagonal nodes) divide the stream of photons into two portions. One portion moves vertically and the other travels diagonally. \textbf{Pass junctions} (circular white nodes) allow the photons to proceed along their initial directions. \textbf{Converge junctions} (circular yellow nodes) play a role in transferring photons from diagonal lines to vertical lines. Though the circular yellow nodes overlap with the hexagonal nodes in the abstract network, they are physically separate, as shown in (a). Photons travelling diagonally from a split junction to the next split junction represents including an element into the summation. The value of the element is equal to the number of junctions between two subsequent rows of split junctions, as denoted by the integers on the left. The generated subset sums are equal to the spatial positions of the output signals, as the port numbers denote. \textbf{(c)} The \emph{X-Y} view of the top left corner of the waveguide network in (a) and the abstract network in (b), is composed of the three basic junctions whose \emph{X-Z} views are shown in \textbf{(d)}, \textbf{(e)} and \textbf{(f)} respectively. The split junction is realized by a modified 3D beam splitter where a coupling distance of $10 \mu m$ , a coupling length of $1.8$ mm and a vertical decoupling distance of $25 \mu m$ are deliberately selected, leading to a desirable splitting ratio. The unbalanced output of split junctions, revealed by the intensity distribution in (d), is designed to compensate the bending loss caused by the subsequent arc $\wideparen{cm}$ and arc $\wideparen{nf}$ in (c). The converge junction is almost a mirror-image split junction except a different coupling length of $3.3$ mm. The residual in port \emph{g} is small enough to be ignored. A vertical decoupling distance of $25 \mu m$ guarantees an excellent pass junction whose extinction ratio is around $24$ dB, as the intensity distribution in (f) presents.}
\label{fig1}
\end{figure*}

Despite the immense difficulty, some researchers attempt to solve NP-complete problems in polynomial time with polynomial resource. A memcomputing machine \cite{Traversa_2015_IEEE,Ventra_2013_NatPhys} as powerful as a NTM has been demonstrated while the ambitious claim is not valid in a realistic environment with inevitable noise \cite{Traversa_2015_SA}. Designs of a NTM, where the magical oracles \cite{Church_1937_logic,Dawson_2007_ModLog,Garey_1979_computers} are realized by simultaneous exploring all computation paths, are proposed \cite{Currin_2017_interface,Dolev_2006_USpatent}. Though in the cost of space or material, parallel exploration provides an alternative to decrease time consumption. As time is irreversible, not reusable and completely out of our charge, it is reasonable to trade physical resources for it. Besides the above NTM proposals, similar measurements have been taken, for instance, the increasingly powerful electronic supercomputers with an integration of an increasing number of processors \cite{top500}, molecule-based computation utilizing large quantities of DNAs or motor molecules \cite{Aoi_1998_JPappliedPhys,Henkel_2007_Biosystems,Nicolau_2016_PNAS,Heldt_2018_conference,Delf_2018_focus}. Furthermore, optimized algorithms are applied to specific instances \cite{Horowitz_1974_ACM,Pisinger_1999_Algori,Koiliaris_2017_symposium}.

Though improvements have been made, conventional electronic computers are ultimately limited by heat dissipation problem \cite{Nicolau_2016_PNAS} which is also a possible limitation for memcomputing machines consisting of commercial electronic devices \cite{Traversa_2015_SA}. The molecule-based computation is limited by the slow movement\cite{Nicolau_2016_PNAS,Heldt_2018_conference,Delf_2018_focus} or the long reaction process\cite{Aoi_1998_JPappliedPhys,Henkel_2007_Biosystems}. Quantum computation is still hindered by decoherence and scalability \cite{Chang_2009_conference,Ladd_2010_Nature}. Other proposals are still in the stage of theory \cite{Currin_2017_interface,Dolev_2006_USpatent,Oltean_2009_NatComputing,Hasan_2011_NatComputing,Rudi_2013_OpticsLaser}. However, we notice that photons have been extensively applied in proof-in-principle demonstrations of supercomputing \cite{Caulfield_2010_NatPhoton} even without quantum speed-up, including NP problem such as prime factorization \cite{Nitta_2009_JPappliedPhys} and NP-complete problems such as travelling salesman problem \cite{Shaked_2007_ApplOpt}, Hamiltonian path problem \cite{Wu_2014_Light,Vazquez_2018_OE,Dolev_2010_TheoCS} and dominating set problem \cite{Goliaei_2012_ApplOpt}. The \#P-complete problem, boson sampling \cite{Tillmann_2013_NatPhoton, Spring_2013_Science,Broome_2013_Science,Crespi_2013_NatPhoton,Carolan_2014_NatPhoton,Carolan_2015_Science}, and other computational functions \cite{Feldmann_2017_NatCom}, algorithms \cite{Tang_2018_SA,Tang_2018_NatPhoton} are also demonstrated in a photonic regime. The successful applications imply that photons are potential excellent candidates to solve the SSP.

Here we present a photonic computer constructed with chips serving as processing units to solve the SSP in a physically scalable fashion. Like the current signal in electronic computers or the molecule in molecular computers, photons contained in the optical source are treated as individual computation carriers. They travel in chips along buried waveguide networks to perform parallel computations. The specific instances of the problem are successfully encoded into the networks according to particular rules. The existence of target sums are judged by the arrival of photons to the corresponding output ports of the networks. We further investigate its scalability and performance in time consumption, showing the photon-enabled advantages.

\section*{\large Results}

\subsection{Configuration of the photonic computer for the SSP}

The proposed photonic computer solving the SSP can be classified as a non-Von Neumann architecture, see Supplementary Materials for its role in the evolution of computers. As shown in Fig. \hyperref[fig1]{1(a)}, the photonic computer consists of an input unit, a processing unit and an output unit. The input unit is employed to generate horizontally polarized photons at 810 nm. Photons are then coupled into the processing unit to dissipate into the waveguide network to execute the computation task. After photons emit from the processing unit, the evolution results are read out by the output unit. Here the processing unit is an analog to the CPU of an electronic computer, playing a key role in the computation. In the following, we will discuss the design of the processing unit from mathematical and physical-implementation aspects to illuminate its capability of solving the SSP.

The processing unit can be represented by an abstract network composed of nodes and lines, which is primarily based on the proposal of Dan V. Nicolau Jr. et al \cite{Nicolau_2016_PNAS} while physical implementation has to be designed to fit integrated photonics. As the network for the specific instance $\{2, 5 ,7, 9\}$ in Fig. \hyperref[fig1]{1(b)} shows, there are three different types of nodes representing split junctions, pass junctions and converge junctions, respectively. It should be noticed that though the circular yellow nodes overlap with the hexagonal nodes in the abstract network, they are physically separate, as the waveguide network in Fig. \hyperref[fig1]{1(a)} presents. Once the photons enter the network from the top node, the computation process is activated. The photons are split into two portions at hexagonal nodes (\emph{split junctions}), traveling vertically and diagonally. When meeting the circular white nodes (\emph{pass junctions}), the photons proceed along the original directions. Meanwhile, the circular yellow nodes (\emph{converge junction}), located at the end of the diagonal routes which start from the former row of hexagonal nodes, are responsible for transferring photons from diagonal lines to vertical lines before next splits.

\begin{figure*}
\centering
\includegraphics[width=2 \columnwidth]{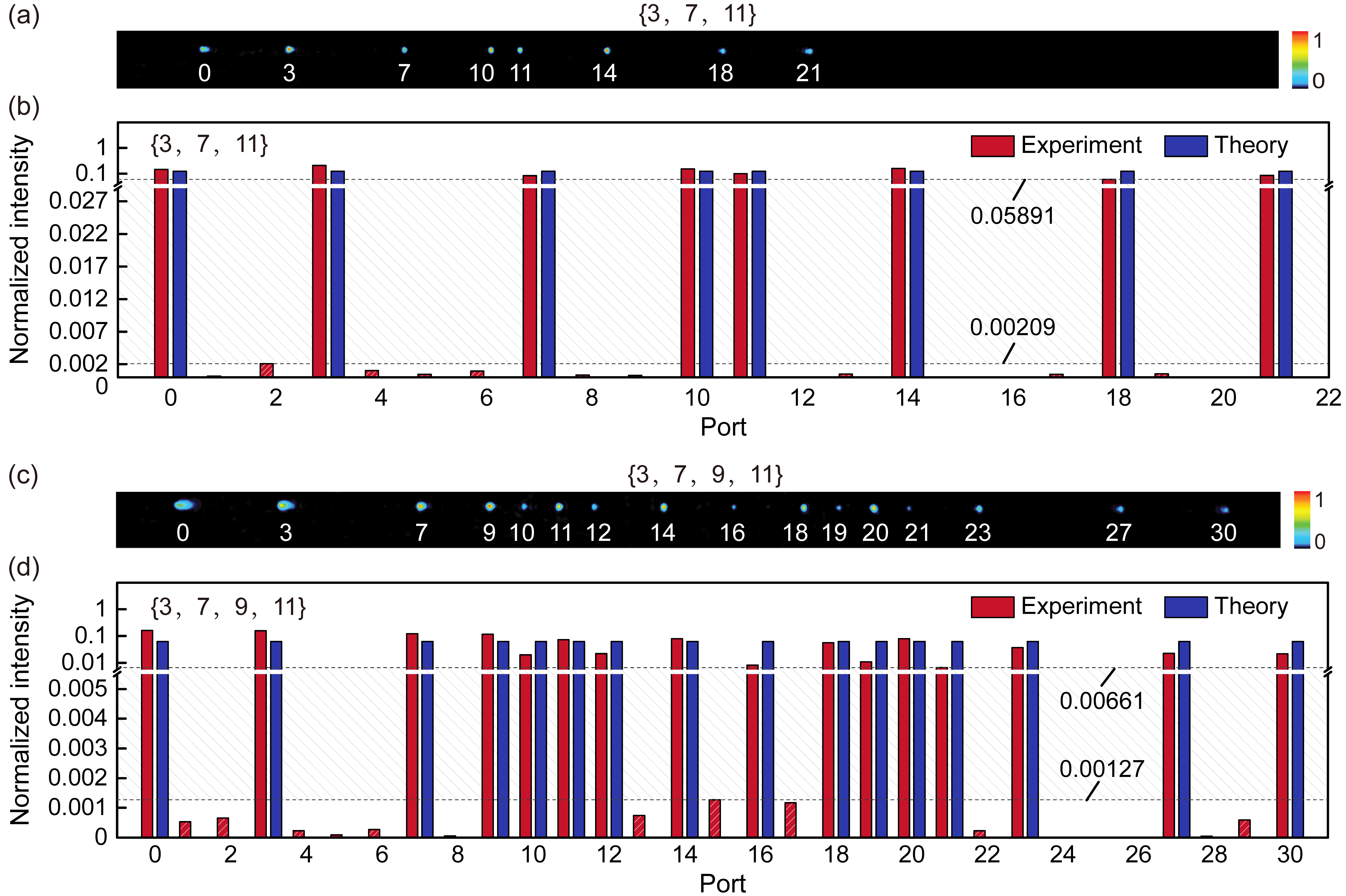}
\caption{\textbf{Experimental read-out of computing results.} \textbf{(a)} Experimental read-out of evolution results of the case $\{3,7,11\}$. Every observable spot certifies the existence of the subset sum denoted by the integer below. \textbf{(b)} Normalized intensity distribution of the case $\{3,7,11\}$ in experiment and theory. Here an axis break is applied to display data points with a value of zero and the logarithmic coordinate simultaneously. The theoretical results are either zero or 0.125 while the experimental results have a fluctuant distribution. A reasonable threshold can be easily found to classify the experimental outcomes into appearance (beyond the threshold) and absence (below the threshold, which is highlighted with slash filling pattern). A wide tolerance band (filled by slash) allows a wide range of threshold with a lower bound of 0.00209 and an upper bound of 0.05891. \textbf{(c)} Experimental read-out of evolution results of the case $\{3,7, 9, 11\}$.  \textbf{(d)} Normalized intensity distribution of the case $\{3,7,9,11\}$ in experiment and theory. Theoretical results are either zero or 0.0625. A wide tolerance band (filled by slash) allows a wide range of threshold with a lower bound of 0.00127 and an upper bound of 0.00661.}
\label{fig2}
\end{figure*}

The specific SSP is encoded into the network according to particular arithmetical and scalable rules: \emph{(i)} The vertical distance (measured as the number of nodes) between two subsequent rows of hexagonal nodes is equal to the value of the element from the set $\{2, 5 ,7, 9\}$, as denoted by the integers on the left. \emph{(ii)} The diagonal routing leads to a horizontal displacement of photons, whose magnitude is also equal to the integer on the left. The diagonal movement of photons represents that the corresponding element is included into the summation. On the contrary, the vertical movement means the element is excluded from the summation. \emph{(iii)} The value of the ultimate sums are equivalent to the spatial position of the output signals, as denoted by the port numbers. For example, the path for port $14$, highlighted by a translucent gray band, reveals that only elements $5$ and $9$ contribute to the subset sum $14$. Owing to the vast parallelism, the photons arrive at the output ports with all possible subset sums generated.

We fabricate the processing unit in Corning Eagle XG glass with femtosecond laser direct writing technique (see Materials and Methods). The top left corner of the waveguide network in Fig. \hyperref[fig1]{1(a)} and the abstract network in Fig. \hyperref[fig1]{1(b)}, is detailedly depicted in Fig. \hyperref[fig1]{1(c)}-\hyperref[fig1]{1(f)}. As we can see, the split junction is realized by a modified 3D beam splitter where the two waveguides first couple evanescently (red segment), and then decouple with one of the waveguides climbing upward and the other proceeding along the initial direction. To avoid extra loss, a vertical decoupling distance of 25 $\rm{{\mu}m}$ is deliberately selected. Meanwhile, the coupling length and coupling distance is set to be 1.8 mm and 10 $\rm{{\mu}m}$ respectively to achieve the desirable splitting ratio. As the intensity distribution in Fig. \hyperref[fig1]{1(d)} reveals, the modified beam splitter is unbalanced, which is on the purpose of compensating the bending loss caused by the subsequent arc $\wideparen{cm}$ and arc $\wideparen{nf}$.

The converge junction is almost the mirror-image split junction except a different coupling length of 3.3 mm. The photons in path $fg$ should be completely transferred to path $eh$ in an ideal case, and the residual induced by the imperfect fabrication is minimized. The output intensity at port $g$ is less than $3\%$ of that at port $h$. The pass junction is implemented in the form of one waveguide crossing over the other at a decoupling vertical distance of 25 $\rm{{\mu}m}$. As the output intensity in Fig. \hyperref[fig1]{1(f)} reveals, the three-dimensional architecture ensures an excellent pass junction whose extinction ratio is around 24 dB. Supported by the powerful fabrication capability of femtosecond laser writing technique \cite{Szameit_2007_OE,Osellame_2012_book}, we are able to map the abstract network of the SSP into a three-dimensional photonic chip.\\

\begin{figure}
\centering
\includegraphics[width=1\columnwidth]{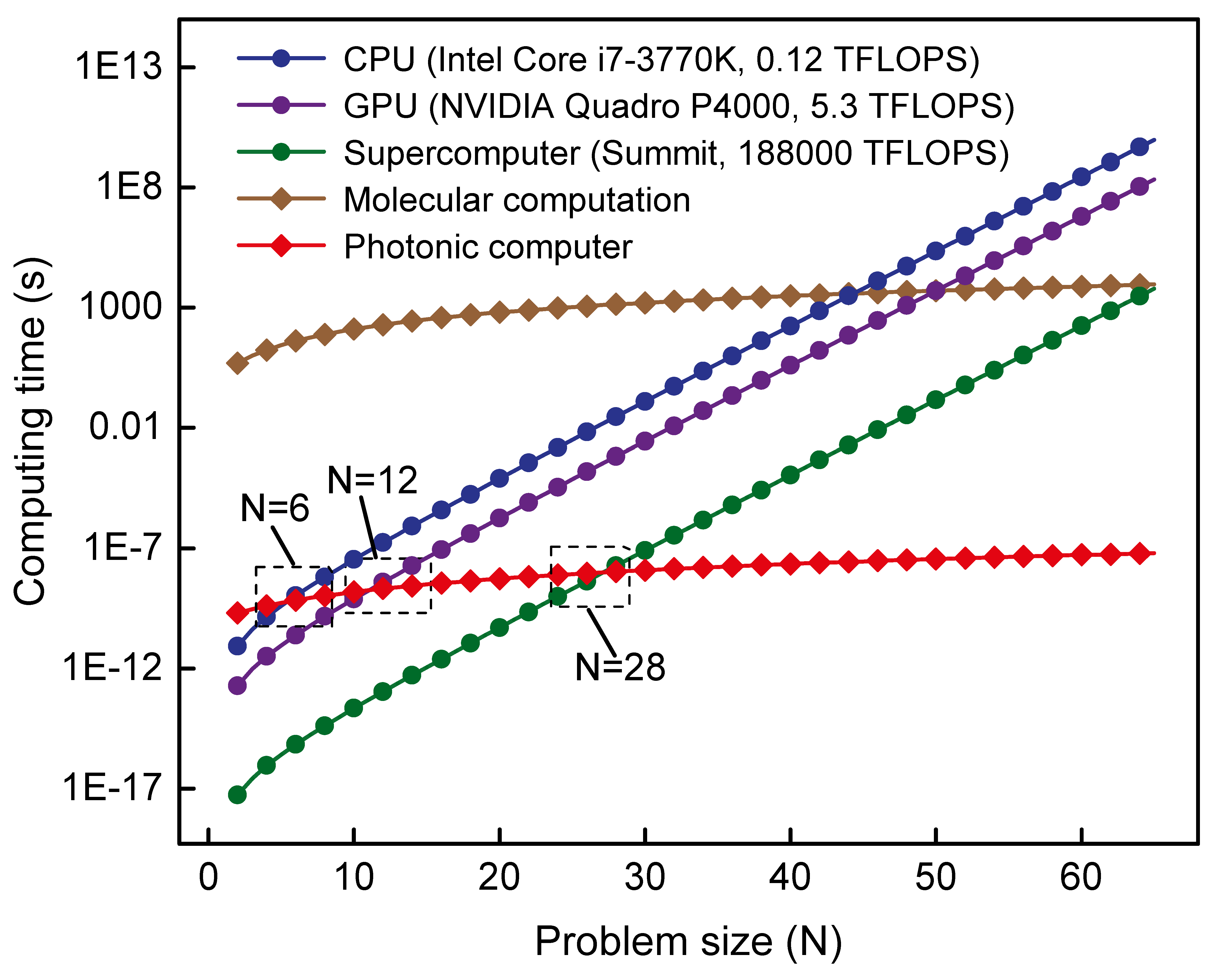}
\caption{\textbf{Time consumption performance.} The comparison of estimated computing time between the photonic computer and other competitors in the case of successive primes $\{2,3,5,7,\dots\}$. The molecular computation is beat by the photonic computer all the time. The electronic competitors working in a brute-force manner are surpassed at $N=6$, $N=12$ and $N=28$, respectively. As problem size increases, the superiority of photonic computer is enhanced, with the computing time of several orders of magnitude shorter than the rivals.}
\label{fig3}
\end{figure}
\subsection{Experimental demonstration}

We demonstrate the computation of the SSP at the specific cases of $\{3,7, 11\}$ and $\{3,7,9, 11\}$. As is shown in Fig. \hyperref[fig2]{2(a)} and Fig. \hyperref[fig2]{2(c)}, the evolution results are read out from a one-shot image, where the photons appear in a line of spots. Every single spot is an accepted witness of the existence of the corresponding sum (denoted by the integer below the spot) if the experiments are trusted. Since the involved problem size is not too large, we check the reliability of our experimental results by enumeration and conclude that all the spots observed are supposed to appear and that none of the expected results is absent.

The reliability of our experiments is further investigated by a comprehensive analysis of the intensity distribution, as presented in Fig. \hyperref[fig2]{2(b)} and Fig. \hyperref[fig2]{2(d)}. We calculate the theoretical distribution through a lossless model consisting of balanced split junctions, perfect pass junctions and ideal converge junctions. Therefore, the theoretical outcomes can be regard as benchmarks of the SSPs. For the case of $\{3,7, 11\}$ , the theoretical result is either zero or $0.125$, while it is either zero or $0.0625$ in the case of $\{3,7,9, 11\}$. In this theoretical regime, zero intensity indicates that a sum does not exist, otherwise it exists.

We apply a threshold to analyze the retrieved intensity for every output port. A valid appearance can be identified if the intensity goes beyond a reasonable threshold, otherwise an absence can be confirmed (highlighted by slash pattern). The tolerant intervals of the thresholds applicable in our experiment are presented with bands filled with slash in Fig. \hyperref[fig2]{2(b)} and Fig. \hyperref[fig2]{2(d)}, straightforward revealing the lower bounds and the upper bounds. Beneficial from the good signal-to-noise ratio obtained in our experiments, there is a wide tolerant band to accept a large range of thresholds, which implies the great accuracies of our experiments and verifies the feasibility of our approach.\\

\subsection{Time-consumption budget}

Interestingly, we find that the optical source launched into the photonic circuit has a significant influence on the performance of our photonic computer. It should be noticed that the photonic supremacy in time consumption over other schemes is achieved by classical light (a stream of photons), not quantum light. We obtain the same evolution results with both classical light and quantum light, and the heralded single-photon source fails to outperform the classical light. This phenomenon is attributed to the fact that a bunch of photons arrive together in the case of classical light while heralded single-photon source only launches one photon at a time. Under such circumstances, it takes longer time with quantum light to collect enough signal photons to be distinguished from the environment, leading to a worse performance than classical light and making it more challenging to surpass electronic computers (see Supplementary Materials).

To show the photon-enabled advantages, we further investigate the time-consumption performance in the case of classical light. Here the computing time is determined by the propagation speed of photons and the longest path in the waveguide network. Owing to the fast movement of flying photons and the compactness of the chip-based networks, it only takes the processing units a fraction of one nanosecond to accomplish the computations in our experiments, which has already surpassed many representative electronic computers emerging in these decades (see Supplementary Materials).

Furthermore, the potential of our approach is explored in the context of successive primes by comparing with other competitors (see Materials and Methods for time estimation of different approaches), as shown in Fig. \hyperref[fig2]{3}. It is noticed that the photonic computer has a significant advantage over the molecular computation, which is attributed to the similar time scaling resulting from the similar configuration of the computing networks, and the superiority of photons in moving speed over molecules, i.e., $\scriptsize{\sim}2\times10^{11}$ mm/s for $810$ nm photons in waveguides and $\scriptsize{\sim}5\times10^{-3}$ mm/s for actin filaments \cite{Nicolau_2016_PNAS}. Though faster biological molecules are reported in a latest research \cite{Delf_2018_focus}, they are still on the long journey of chasing after photons.

The time consumption of representative electronic competitors with the conventional Von Neumann architecture, characterized by floating point operations per second (FLOPS) \cite{flops}, are also presented. It is found that the photonic computer outperforms the state-of-the-art CPU \cite{IntelCore_2012} at a small size that is probably accessible in subsequent experimental demonstrations. Compared with the GPU \cite{Nvidia}, the photonic computer exceeds it until $N=12$. Apparently, it is increasingly challenging to beat an increasingly strong competitor. Nevertheless, the most powerful supercomputer \cite{top500}, Summit, composed of an enormous number of CPUs and GPUs, can be also surpassed at a modest size of $28$. Besides, the superiority of the photonic computer is reinforced with the growth of problem size, as the trend reveals. Even at a medium size, our approach consumes many orders of magnitude shorter computing time than the molecular and electronic rivals, exhibiting strong competitiveness in solving the SSP at the case of successive primes (see Materials and Methods for the speed-up of our photonic computer).\\

\section*{\large Discussion}
In summary, we demonstrate a photonic computer solving the SSP by mapping the problem into a waveguide network in a three-dimensional architecture. With the demonstrated standardized structure of basic junctions, regular configuration of the network and the mature femtosecond laser writing technique, the SSP can be encoded into a physical network and conveniently solved in a scalable fashion. The computational power is further analyzed by investigating the time-consumption performance. The results suggest that, for successive primes, photonic computers are very likely to beat the most powerful supercomputer with a near-future accessible problem size. Other performances, such as signal-to-noise ratio and Fisher information are also discussed (see Supplementary Materials).

The photon-enabled advantage in solving the SSP can be understood from the unique features of light. Firstly, light is essentially a stream of photons, which can be sufficiently employed to probe all the paths in parallel, by being dissipated into a large network with very small fraction of light in each path (can be down to single-photon level). Secondly, the ultimate speed of flying photons makes the evolution time very short in the designed structures, even for a large and complicated photonic network. Thirdly, photons can be confined in a very limited space with the technique of integrated photonics, which is beneficial to both the computing speed and scalability. Last but not least, interference is a unique strength of photons, whereas we can not see its contribution to the speed-up of the proposed photonic computer. Nevertheless, it can be potentially utilized to achieve a reconfigurable photonic computer for different SSPs in the future (see Supplementary Materials).

Besides the fundamental interest of racing with conventional electronic computers, it would be more fascinating to map many real-life problems into the frame of solving the SSP, which may boost the building of such photonic computer towards industrialization. It is also possible, but still open, to solve other NP problems in this purpose-built photonic computer. In light of the fact that any NP problem can be reduced to a NP-complete problem efficiently \cite{Karp_1972_complexity}, any NP problem is able to be mapped to the proposed network in principle. Therefore, a photonic solution of the SSP implies possible solutions of a wide range of NP problems. Moreover, photon-enabled unique feature may also show its strength in other new computing architectures \cite{Smolyaninov_2019_JOSA,Caligiuri_2019_JPCS}.\\

\section*{\large Materials and Methods}
\noindent\textbf{Photonic chips fabrication.}
Waveguide networks in three-dimensional artchitecture are written by the femtosecond laser with a repetition rate of $1$ MHz, a central wavelength of $513$ nm, a pulse duration of $290$ fs and a pulse energy of $190$ nJ. Before radiating into the borosilicate substrate at a depth of $170 \mu m$, the laser beam is shaped by a cylindrical lens and then focused by a $100\times$ objective with a numerical aperture of $0.7$. During the fabrication, the translational stage moves in $X$, $Y$, $Z$ directions according to the user-defined programme at a constant speed of $15$ mm/s. The careful measurements and characterization on the geometric parameter dependence of the three types of junction, such as coupling length, coupling distance, decoupling distance and curvature, are taken to optimize the performance to form the standard elements.\\

\noindent\textbf{Estimation of computing time.}
For both molecular computation and our approach, the computing time is determined by the moving speed of computation carrier (i.e., molecules and photons) and the longest path in the network. For example, in the case of $\{2,5,7,9\}$ shown in Fig. \hyperref[fig1]{1}, the longest path is the one linking to the port 23 which represents the sum of all the elements in the set. According to the geometrical parameters and the scalable rules of our waveguide network, it is easy to calculate the length of the longest path. The propagating speed of photons is estimated on the basis of the refractive index of Corning Eagle XG \cite{refractiveindex} and the refractive index change induced by femtosecond laser writing \cite{Eaton_2008_OE}. The structural parameters of molecular computation derive from the experiment by Dan V. Nicolau Jr. et al \cite{Nicolau_2016_PNAS}. The faster molecules, actin filaments, are chosen to compared with our approach.

The running time taken by conventional electronic computers working in a brute-force mode, searching the entire solution space consisting of all possible subsets, to solve the SSP is estimated by multiplying FLOPS by the total number of arithmetic operations. The data of FLOPS adopted in our research is either the peak performance or theoretical performance of the corresponding electronic machine. Performance degradation \cite{IntelCore_2012} in practical scenario is neglected.\\

\noindent\textbf{Speed-up of the photonic computer.}
Given a set of $N$ elements, the number of subsets grows exponentially with $N$. According to the definition of the SSP, it requires us to verify every possible subset. If we regard the verification of a subset as a subtask, the number of subtasks or the number of computation operations increases at an exponential rate.

For a conventional electronic computer working sequentially, all subtasks are executed in sequence. Therefore, the total computing time is equivalent to the product of the number of computation operations and the unit time taken by a single operation, growing at an exponential rate. For our photonic computer which works in a parallel mode, all subtasks can be executed simultaneously. In our implementations, each subset is mapped to a path of the photonic circuits. With light beam (a stream of photons) being split and propagating along all possible paths, all subsets are verified at the same time. On such an occasion, the total computing time only depends on the verification of the largest subset.

Here the verification of the largest subset corresponds to the movement of photons from the input port to the output port through the longest path. As a result, the computing time is equal to the traveling time of photons in the longest path, growing at a sub-exponential rate which is slower than that of electronic computers. Moreover, as photons possess an ultra-high propagating speed and the integrated photonic circuit has a compact structure, the computing process is further speeded up.\\

\section*{\large Supplementary Materials}
\noindent{I. The evolution of computers and the role of non-Von Neumann architecture}

\noindent{II. The influence of optical source on time consumption}

\noindent{III. Time-consumption performance}

\noindent{IV. Signal-to-noise ratio}

\noindent{V. Fisher information}

\noindent{VI. The role of interference}

\noindent{FIG. S1: The role of non-Von Neumann architecture.}

\noindent{FIG. S2: Time-consumption performance.}

\noindent{FIG. S3: Signal-to-noise ratio.}

\section*{\large Acknowledgments}
\textbf{Funding:} This research is supported by the National Key R\&D Program of China (2019YFA0308700, 2017YFA0303700), the National Natural Science Foundation of China (61734005, 11761141014, 11690033, 11774222), the Science and Technology Commission of Shanghai Municipality (17JC1400403), the Shanghai Municipal Education Commission (2017-01-07-00-02-E00049). X.-M.J. acknowledges additional support from a Shanghai talent program. \textbf{Author contributions:} X.-M.J. and H.-P.Z. conceived the project. X.-M.J. supervised the project. X.-Y.X. and X.-M.J. designed the experiment. X.-Y.X. fabricated the photonic chips. X.-Y.X., X.-L.H., Z.-M.L., J.G., Z.-Q.J., Y.W., R.-J.R. and X.-M.J. performed the experiment and analyzed the data. X.-Y.X. and X.-M.J. wrote the paper with input from all the other authors. \textbf{Competing interests:} The authors declare that they have no competing interests. \textbf{Data and materials availability:} All data needed to evaluate the conclusions in the paper are present in the paper and the Supplementary Materials. Additional data available from authors upon request.

\clearpage
\newpage

\onecolumngrid
\section*{\large Supplemental Materials: A Scalable Photonic Computer Solving the Subset Sum Problem}
\setcounter{figure}{0}
\setcounter{table}{0}
\setcounter{equation}{0}
\renewcommand{\figurename}{Supplementary Figure}
\renewcommand{\tablename}{Supplementary Table}

\renewcommand{\thetable}{\arabic{table}}
\renewcommand{\theequation}{{S}\arabic{equation}}

\bigskip
\section*{\large Supplementary Note 1: The evolution of computers and the role of non-Von Neumann architecture}
The revolution brought by advanced electronic computers is prevalent in modern life, involving education, finance, transportation, scientific research, medical care and so on. However, it is hard to imagine such scenarios take place in the past days when people calculated with manual abacus or electromechanical devices. It is the development of computing devices that improves the human computing power, leading to a higher working efficiency and then propelling the prosperity of society.

Looking back into the history of electronic computers, it evolves from a crude model to a sophisticated machine. The first general-purpose electronic computer was built with vacuum tubes in 1950s, weighing 30 tons. Then it developed with the emergence of transistors and thrived with the innovation of integrated circuits in the past decades, following Moore's law towards higher performances and more compact size, boosting the improvement of human computing power. Recent electronic computers enable performances comparable to a mouse brain, far beyond the levels of early ones, such as UNIVAC.

\begin{figure*}[ht]
\centering
\includegraphics[width=1\columnwidth]{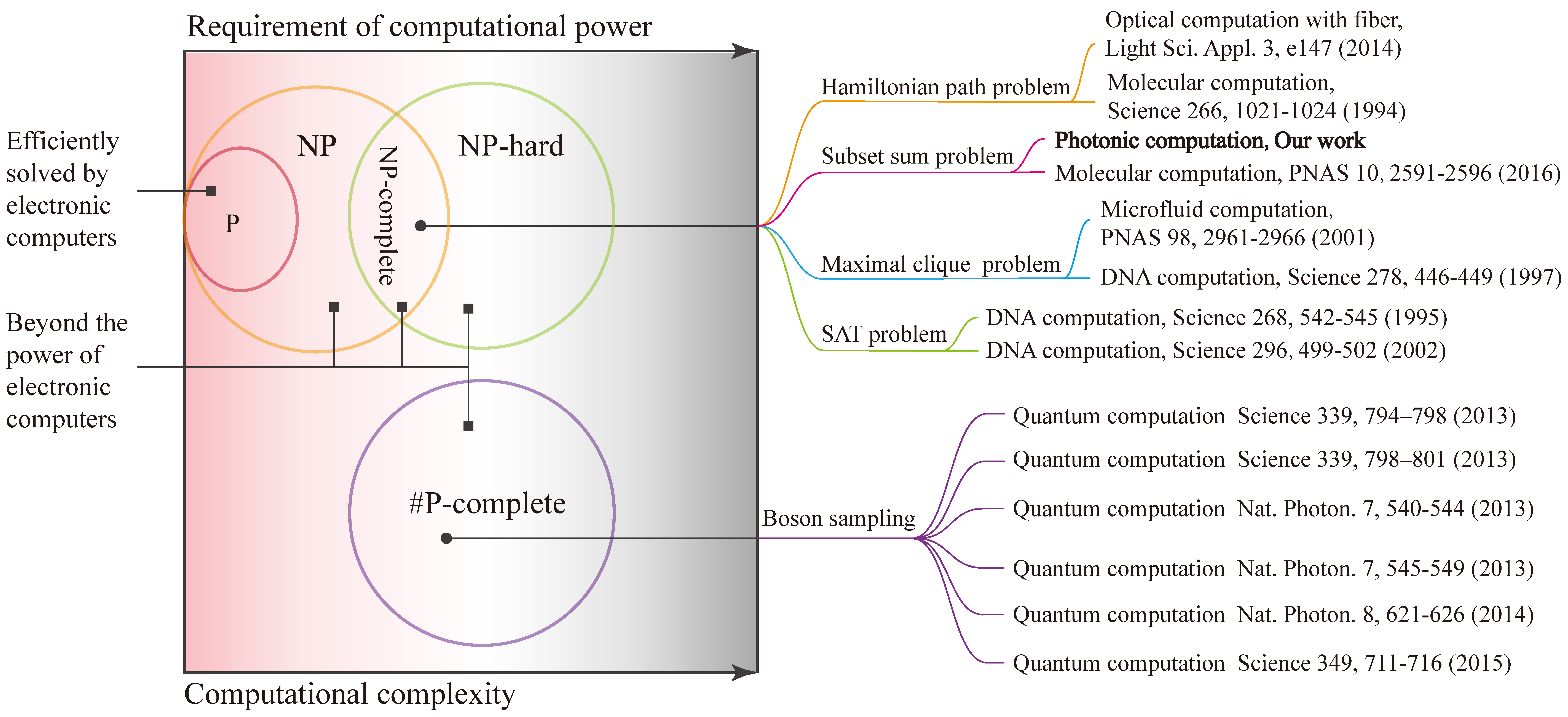}
\caption{\textbf{The role of non-Von Neumann architecture.} Various non-Von Neumann computing architectures have been successfully applied to solve the problems intractable for the conventional Von Neumann architecture.}
\label{fig1}
\end{figure*}

However, Moore's law has already faltered in recent years and it is believed to go to an end in the near future. Due to the inevitable heat dissipation problem, traditional electronic computers are ultimately limited. Meanwhile, with the miniaturization of integrated circuit, transistors will be unreliable as quantum tunneling effect is emerging. The human computing power counting on electronic computers is becoming increasingly difficult to make a breakthrough. Up to today, even the most powerful supercomputer operates an order of magnitude slower than the human brain, suggesting that the ultimate limit is hardly dramatically beyond the human brain.

Furthermore, conventional electronic computers based on Von Neumann architecture are inherently not powerful enough to solve NP-complete problems efficiently. As we depict in FIG. S1, many combinatorial problems in practice can be classified into corresponding categories according to their computational complexity. Problems belonging to the subset P can be efficiently solved by conventional computers while other NP problems and \#P problems can't be solved in polynomial time. The intractability is attributed to the fact that electronic computers work sequentially while the computing workload grows exponentially or super-exponentially with the increase of problem size, which leads to a higher requirement of computational power.

To meet the present challenges and to embrace a future with infinite possibilities brought by the revolution of human computing power (like the revolution induced by electronic computers decades ago), looking for novel computing architectures with a potential to improve the limit of human computing power is of great significance. Over these years, several novel computing architectures independent of Von Neumann architecture have been paid high attention, including quantum computation, optical computation, DNA computation, molecular computation and microfluid computation. As listed in the right of FIG. S1, specific NP-complete problems or \#P-complete problems are regarded as benchmarks to evaluate the performances of these computing architectures, since the power required to efficiently solve them is far beyond the capability of conventional computers. Most novel computing architectures are first demonstrated for specific purposes, whereas, like the early electronic computers, they provide potentially feasible accesses to the breakthrough of human computing power.

Being in the age of exploration and discovery, we believe that it is non-trivial to make attempts in various computing architectures. Photonic computation can be regarded as a promising candidate, as photons possess unique features enabling higher computing power, such as high propagating speed for fast computation, ultra-low energy for energy-efficient computation and the ability of being split for parallel computation. In our research, we have proposed a scalable photonic computer where abundant photons are treated as individual computation carriers, moving at high speed in the integrated photonic circuit embedded in silica chips, to perform computations in parallel. Analogous to the specific-purpose quantum computer for boson sampling, an intractable \#P problem, the strengths and potential of the photonic computer are demonstrated by solving the subset sum problem (SSP), a typical NP-complete problem intractable to electronic computers.

Compared with other computing architectures, such as electronic computers and molecular computation, the proposed photonic computer outperforms them and shows supremacy in time consumption with many orders of magnitude less computing time. Unlike conventional electronic computers whose working mode is sequential, i.e., operations are executed one by one, the photonic computer allows a parallel working mode that is more powerful to handle the heavy workload of solving the SSP. In contrast to the slow movement of motor protein utilized in molecular computation, the photons applied in our protocol move at an ultra-high speed (equivalent to operating extremely fast), which is more beneficial to save time. Furthermore, distinguished from other optical schemes, like Fourier transformation and fiber computing networks which are hard to scale up, our chip built-in photonic computer is scalable, laying a strong foundation for the further investigation of large-size problems.

\section*{\large Supplementary Note 2: The influence of optical source on time consumption}
We have applied a coherent laser to launch photons into the photonic computer to solved the SSP. As Fig. 3 in the main text exhibits, the photonic computer shows a supremacy over other competitors in time consumption with the growth of problem size.

Given the quantum supremacy taking place in other situations such as boson sampling (Ref. 34-39) and quantum walk (Ref. 41), we attempt to figure out the influence induced by quantum light on the time consumption of solving the SSP with our photonic computer. Here we consider a heralded single-photon source. It is found that the same evolution results are obtained with both classical light and quantum light, and the heralded single-photon source fails to outperform the classical light. This phenomenon is attributed to the fact that a bunch of photons arrive together in the case of classical light while heralded single-photon source only launches one photon at a time. Under such circumstances, it takes longer time with quantum light to accumulate enough signal photons to be distinguished from the environment, resulting in a worse performance and making it more challenging to surpass electronic computers.

Assuming that the environment noise is equivalent to $m$ photons, then it is essential to accumulate at least $m+1$ signal photons to be distinguished from the environment. In an lossless case, the computing time taken by quantum light is at least $m$ times of that taken by classical light. In a practical case with inevitable loss, the computing time taken by quantum light is further increased.

\section*{\large Supplementary Note 3: Time-consumption performance}
Besides the comparison in time consumption displayed in the Fig. 3 in the main text, we further compare the photonic computer with representative electronic computers which are developed over these decades and once treated as the state-of-the-art computers. As FIG. S2 reveals, our first photonic demonstration (N=4) is comparable to today's electronic computer (Intel Core i7-3770k), and indeed outperforms the rest representative electronic computers emerging in the first several decades since the electronic computer was invented in 1950s. Furthermore, the photonic computer has a lower growing rate of computing time than all the electronic counterparts, which dominantly determines the ultimate winner of this race. It has taken conventional computers over half century to evolve from a crude model to state-of-the-art machines, undergoing a development of several generations. We believe that, to some degree, the performance of our first implementation reasonably exhibits the strengths and potential of the photonic computer.

\begin{figure}[ht]
\centering
\includegraphics[width=0.5\columnwidth]{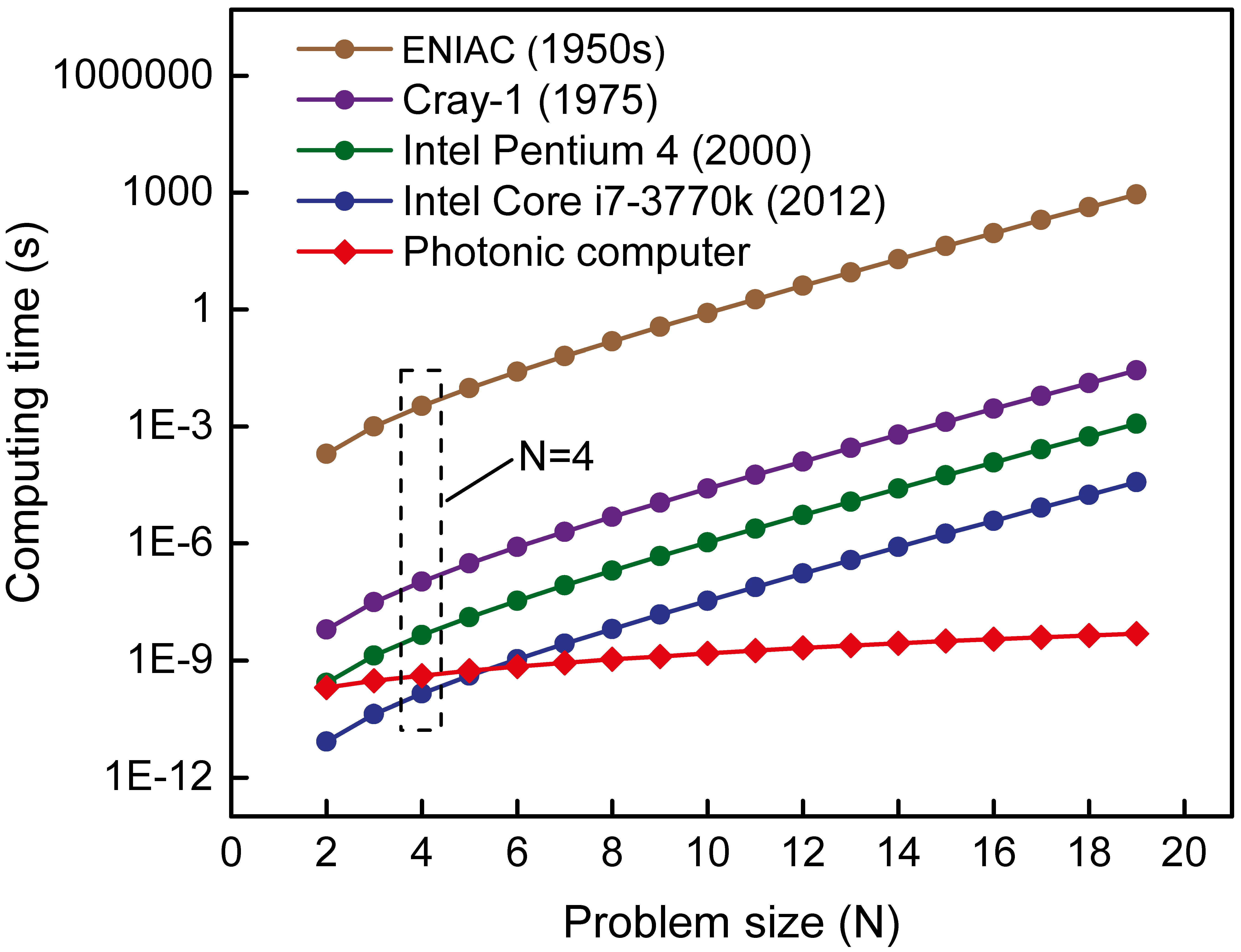}
\caption{\textbf{Time-consumption performance.}The comparison in time-consumption performance between our photonic computer and the representative electronic computers which are developed over these years. These electronic computers are first launched in 1950s, 1975, 2000 and 2012, respectively.}
\label{fig2}
\end{figure}

\section*{\large Supplementary Note 4: Signal-to-noise ratio}
In the proposed photonic computer, the output signal going through the longest path appears be the weakest. Therefore, the signal-to-noise ratio (SNR) is evaluated by analyzing the signal traveling in the longest path.
According to the definition of SNR, we have
$$SNR=10log_{10}(Sig/Noi)$$
$$SNR=-10log_{10}(In/Sig)+10log_{10}(In/Noi)$$
\noindent{Where $Sig$ represents the power of the weakest signal, $Noi$ is the equivalent power of environmental noise and $In$ is the input power.}

We assume a stable environment , i.e. the environmental noise $Noi$ is invariant. The latter term $10log_{10}(In/Noi)$ is independent of problem size $N$ and determines the upper bound of SNR which grows with the increase of $In$. The former term $-10log_{10}(In/Sig)$ is related to problem size $N$. Then the expression can be simplified as
$$SNR=F(N)+C$$
\noindent{Where $C=10log_{10}(In/Noi)$, $F(N)=-10log_{10}(In/Sig)$.}

Based on the design of the photonic computer, we have
$${F(N)=c_{1}N+c_{2}S}$$
\noindent{Where $S$ is the sum of the first $N$ primes and grows at a sub-exponential rate. Constants $c_{1}$ and $c_{2}$ are determined by the specific parameters of the photonic computer, such as propagation loss, splitting ratio of beam splitters and the size of basic modules.} Here $c_{1}$ and $c_{2}$ are estimated to be -3.212 and -0.0252, respectively. Therefore, we have
$$SNR=-3.212N-0.0252S+C$$
Take silicon detector for example, with 100Hz dark count and 1MHz count rate, the noise in one shot is down to 0.0001. The corresponding SNR with different input power is exhibited in FIG. S3, decreasing at a sub-exponential rate. It should be noticed that the upper bound $C$ has no influence on the decreasing rate of SNR.\\

\begin{figure}
\centering
\includegraphics[width=0.5\columnwidth]{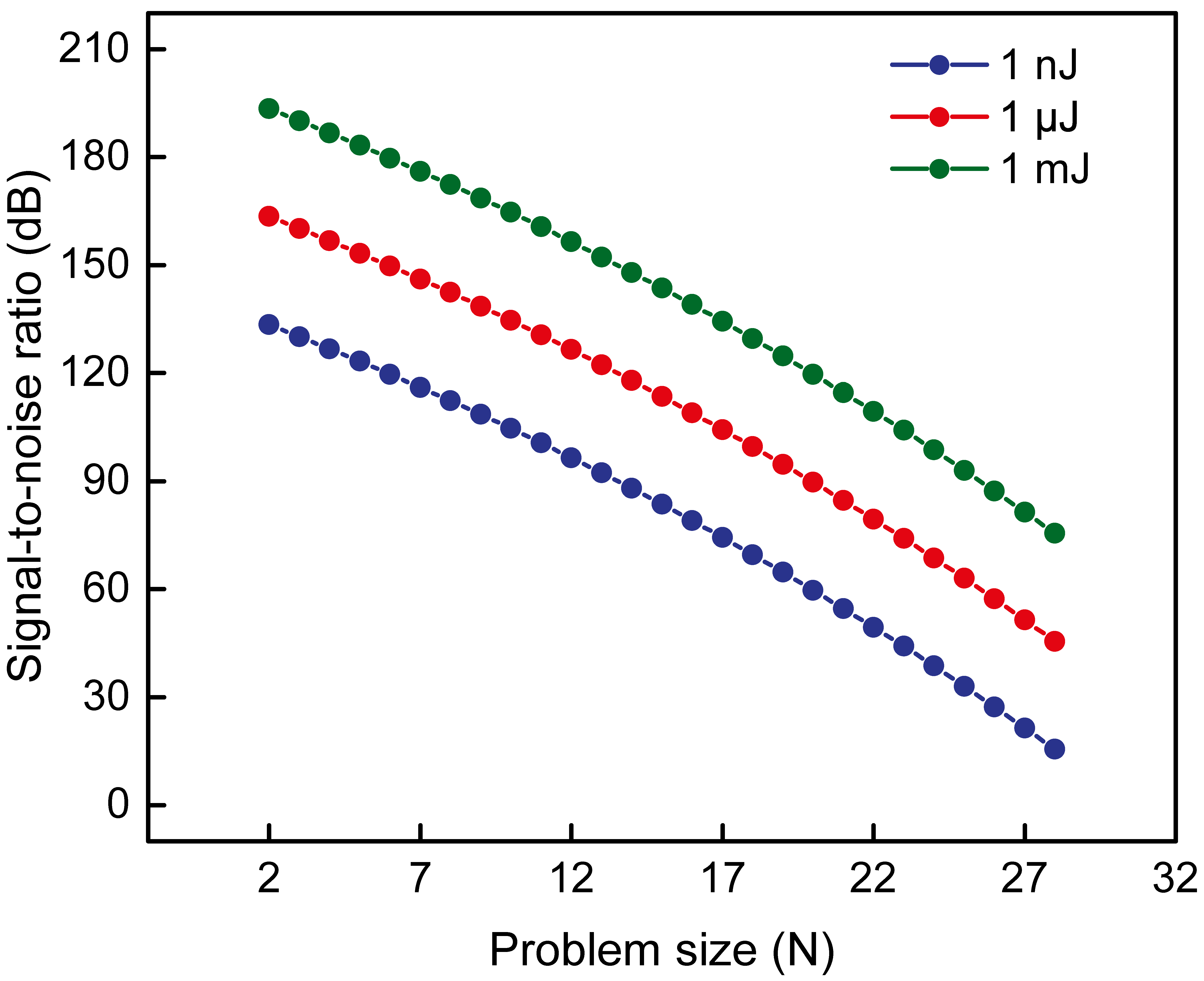}
\caption{\textbf{Signal-to-noise ratio.}The signal-to-noise ratio of our photonic computer in the case of different input power.}
\label{fig3}
\end{figure}

\section*{\large Supplementary Note 5: Fisher information}
Since the longest path suffers the largest loss, the number of trials depends on the probability of a single photon arriving at the corresponding output port of the longest path. On such an occasion, we consider a coin toss model to evaluate the Fisher information, i.e. the variable $X$ only has two outcomes. Here $X$ is a trial. $X=1$ represents that a photon reaches the target output port, which happens with a probability of $\theta$. $X=0$ corresponds the opposite situation, which takes place with a probability of $1-\theta$. Therefore, the probability density function is $$Prob(X|\theta)=\theta^X(1-\theta)^{(1-X)}.$$
\noindent{According to the definition of Fisher information (Ref. 52), the corresponding Fisher information carried by one trial is expressed as} $$Inf(\theta)=\frac{1}{\theta(1-\theta)}.$$ \noindent{Since the Fisher information is additive, the Fisher information contained in $M$ independent trials is $\frac{M}{\theta(1-\theta)}$.} Finally, we acquire the lower bound on the variance of $\theta$ as
$$Var(\theta)\ge\frac{1}{Inf(\theta)}=\frac{\theta(1-\theta)}{M}.$$
\noindent{As $\theta$ is related to problem size $N$, we have}
$$Var(N)\ge\frac{1}{Inf(\theta(N))}=\frac{\theta(N)(1-\theta(N))}{M}.$$
\noindent{Based on the above analysis of SNR, we obtain $\theta(N)=10^{\frac{F(N)}{10}}=10^{\frac{-3.212N-0.0252S}{10}}$,
where $S$ is the sum of the first $N$ primes.}

\section*{\large Supplementary Note 6: The role of interference}
\noindent \textbf{Influence on ultimate outcome.} Based on the structure of the proposed photonic circuit, it is found that different light beams stemming from the same split junctions may gather somewhere else during their travel to the output ports. Therefore, in the case of coherent light, there is a possibility of occurrence of interference, which might induce a fluctuation of the intensity distribution at the output ports.

However, interference can hardly influence the ultimate outcome. Since the SSP is a decision problem whose kernel lies in giving an answer of YES or NO, a fluctuation of intensity distribution almost can't turn the outcome inverse in light of a few facts: \emph{(i)} A high signal-to-noise ratio, up to tens of dB even for a relatively large-size problem, can be provided, as analyzed in the section IV, which lays a strong foundation for tolerating the fluctuation induced by interference. \emph{(ii)} Completely destructive or constructive interference, of course unwanted in our experiment, requires highly precise control of phase, whereas the harsh requirement is of great difficulty to meet due to the inevitable fabrication imperfection. \emph{(iii)} The influence from interference can be greatly weakened and even eliminated by applying broadband optical sources with an ultra-short coherence length, such as superluminescent diodes (e.g., SLD-35-HP SUPERLUM and SLD830S-A10W THORLABS) whose coherence length can be as short as a few microns.\\

\noindent \textbf{Contribution to the speed-up.}
The huge consumption of time of solving the SSP is caused by its inherent intractability that the computation workload grows exponentially with the problem size $N$. Apparently, either a reduction of the time taken by a subtask of the computation workload or applying an advanced working mode to handle the great quantity of subtasks, such as a parallel mode, can speed up the computation process.

In our scheme, the time taken by a subtask is determinate owing to the constant propagation speed of light and the permanent structure of the photonic computer. Besides, the parallel working mode of the photonic computer is achieved by constructing a network composed of parallel computation paths and enabled by the splitting behavior of light based on evanescent coupling. The initial light is split into $2^N$ portions during its travel in the network. These portions act as independent computation carriers to simultaneously execute all the subtasks. Since interference can not produce extra independent computation carriers, we can not see the contribution from it to the speed-up of solving a specific SSP.

However, interference, the unique feature and strength of light, can be potentially utilized to construct a reconfigurable photonic computer for the SSP through routing, which is beneficial to the speed-up of solving different SSPs as the reconfigurable photonic computer can be conveniently converted to the particular configuration for each SSP.\\

\end{document}